\documentclass[aps,twocolumn,showpacs]{revtex4}
\usepackage{amsmath}
\usepackage{graphicx}
\usepackage{dcolumn}
\usepackage{verbatim}
\usepackage{times}
\usepackage{subfigure}
\usepackage{bm}
\usepackage{color}
\usepackage{subfigure}
\usepackage[colorlinks,bookmarks=false,citecolor=blue,linkcolor=red,urlcolor=blue,backref=red,dvipdfm]{hyperref}

\begin{document}

\title{Phase diagram of Kondo-Heisenberg model on honeycomb lattice with geometrical frustration}

\author{Huan Li$^{1}$, Hai-Feng Song$^{2,3}$ and Yu Liu\footnote{E-mail: liu\_yu@iapcm.ac.cn}$^{2,3}$}
\affiliation{$^{1}$College of Science, Guilin University of Technology, Guilin 541004,
China \\
$^{2}$LCP, Institute of Applied Physics and Computational Mathematics, Beijing
100088, China \\
$^{3}$Software Center for High Performance Numerical Simulation, China Academy of Engineering Physics, Beijing 100088, China
}

\date{\today}

\begin{abstract}
We calculated the phase diagram of the Kondo-Heisenberg model on two-dimensional honeycomb lattice with both nearest-neighbor and next-nearest-neighbor antiferromagnetic spin exchanges, to investigate the interplay between RKKY and Kondo interactions at presence of magnetic frustration.
Within a mean-field decoupling technology in slave-fermion representation, we derived the zero-temperature phase diagram as a function of Kondo coupling $J_k$ and frustration strength $Q$.
The geometrical frustration can destroy the magnetic order, driving the original antiferromagnetic (AF) phase to non-magnetic valence bond state (VBS). In addition, we found two distinct VBS. As $J_k$ is increased, a phase transition from AF to Kondo paramagnetic (KP) phase occurs, without the intermediate phase coexisting AF order with Kondo screening found in square lattice systems. In the KP phase, the enhancement of
frustration weakens the Kondo screening effect, resulting in a phase transition from KP to VBS. We also found a process to recover the AF order from VBS by increasing $J_k$ in a wide range of frustration strength.
Our work may provide deeper understanding for the phase transitions in heavy-fermion materials, particularly for those exhibiting triangular frustration.
\end{abstract}

\maketitle

\section{Introduction}

In recent years, the exploration of quantum phase transitions in strong-correlated electron systems continuously attracts attention, among which, the heavy-fermion (HF) systems focus great interests\cite{Si01,Lohneysen07}. For the HF materials, which are theoretically described by Kondo lattice model (KLM) or Kondo-Heisenberg lattice model (KHLM),
the interplay between Ruderman-Kittel-Kasuya-Yosida (RKKY) interaction and Kondo coupling may cause phase transitions between a variety of phases.
For the KLM on two-dimensional square lattice, in addition to the Kondo-destroyed AF phase with small Fermi surface(AF$_s$) in weak Kondo coupling region and the Kondo-screened paramagnetic phase (KP) with large Fermi surface at strong coupling, various works using mean-field approximation, bond-fermion approach, variational Monte Carlo calculations and Gutzwiller approximation have verified a phase coexisting Kondo screening with AF order which exhibits a large Fermi surface (AF$_l$) in intermediate Kondo coupling strength\cite{Zhang00,Watanabe07,Lanata08, Custers12,Isaev13,Eder16}. In this context, the offset between magnetic transition and Kondo breakdown is realized, as observed in a series of materials such as Ce$_3$Pd$_{20}$Si$_6$~\cite{Custers12}, CeCoGe$_{3-x}$Si$_x$~\cite{Duhwa98}, CeRh$_{1-x}$Co$_{x}$In$_{5}$~\cite{Harrison07,Goh08} and YbRh$_{2}$Si$_{2}$ under Co and Ir doping~\cite{Paschen04,Friedemann09}. Notably, the reduction of conduction electron concentration away from half-filling may lead to a offset-to-converge shift between magnetic transition and Kondo breakdown~\cite{Watanabe07,Lanata08,Li15}.

In theoretical models, the RKKY interaction is ordinarily represented by Heisenberg exchanges between local spins. The antiferromagnetic Heisenberg exchanges favor a AF order in the absence of magnetic frustration. while the antiferromagnetic feature of Kondo coupling brings spin fluctuations between local spins and itinerant electrons. The competition between these two mechanisms results in enriched phase diagram. In practice, Isaev~\cite{Isaev13} stated that the increasing of Heisenberg exchanges can generate phase transitions from KP to AF$_l$, then to AF$_s$ phase, successively. Surprisingly, he also found that the variation of next-nearest-neighbor (N.N.N) electron hopping $t'$ can also lead to a offset-to-converge shift between magnetic transition and Kondo breakdown.

While most of previous studies of KHLM focused on the square lattice, the two-dimensional lattice with geometrical frustration provides additional tuning factor into the phase transitions in HF systems, such as CePd$_{1-x}$Ni$_x$Al~\cite{Fritsch14} on Kagome lattice, and YbAgGe~\cite{Tokiwa13,Dong13}, YbAl$_3$C$_3$~\cite{Khalyavin13,Matsumura15} on triangular lattices. For Yb$_2$Pt$_2$Pb~\cite{Kim13,Kim11}, a KHLM on Shastry-Sutherland square lattice (SSL)~\cite{Laeuchli02} was proposed, in which the diagonal Heisenberg exchanges $J_2$ provide geometrical frustration, and the phase diagram was determined on $J_k$-$J_2$ plane within mean-field approximation~\cite{Coleman10,YuRong14}. In addition to the conventional phase transitions between AF$_s$, AF$_l$ and KP phases in weak frustration region, the authors found that the enhanced frustration not only suppresses the AF order and gives rise to a VBS, but can also drive the KP phase to VBS.

In order to investigate the quantum phase transitions in other frustrated HF systems different from SSL, we study the KHLM on honeycomb lattice with N.N.N. exchanges providing magnetic frustration. The local spins are represented by slave-fermions then the phase diagram is calculated through mean-field self-consistent treatments.
We found a distinct phase diagram from the KHLM on SSL. Firstly, no AF$_l$ phase appears; secondly, two different VBS emerge, which are separated by $J_2/J_1$=1; thirdly, we noticed a new process to recover AF order from VBS by enhanced $J_K$ at medium frustration.
Our results may shed light on the explanation of quantum phase transitions observed in HF compounds with geometrical frustration.

\section{Kondo-Heisenberg model with geometrical frustration on honeycomb lattice}
\begin{figure}[tbp]
\hspace{-0.2cm} \includegraphics[totalheight=1.8in]{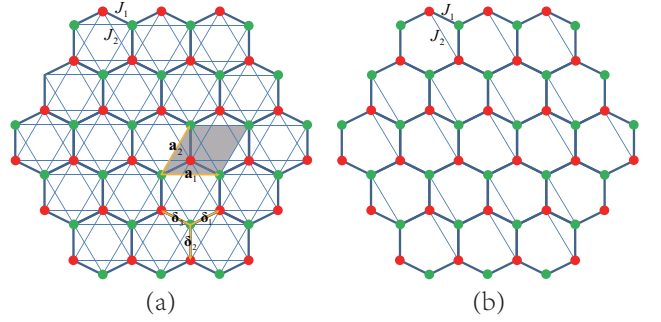}
\caption{(Color online) Kondo-Heisenberg model on honeycomb lattice. Red and green dots denote local spins on two sublattices. $J_1$ and $J_2$ are N.N. and N.N.N. Heisenberg exchanges, respectively. Two cases are considered: six (a) and only one (b) $J_2$-lines exist in each hexagon.
The gray parallelogram in (a) indicates the unit cell. }
\label{lattice}
\end{figure}

We consider the Kondo-Heisenberg model on honeycomb lattice. The lattice structure and RKKY exchanges are shown in Fig.\ref{lattice}(a). The honeycomb lattice can be naturally divided into two sublattices A and B, respectively. Two types of Heisenberg exchanges are considered:
local spins interact with each other along the hexagonal side length direction with strength $J_1>0$; spins also interact with their next-nearest-neighbors with strength $J_2>0$. The nearest-neighbor (N.N.) Heisenberg interactions $J_1$ favor a AF configuration with antiparallel moments on different sublattices, while the N.N.N. interactions $J_2$ provide magnetic frustration to create a non-magnetic VBS. The model Hamiltonian is written explicitly as
\begin{align}
\mathcal{H}=&\sum_{i,j,\sigma}t_{ij}(c^{\dag}_{i\sigma} c_{j\sigma}+h.c.)\nonumber\\
+&\sum_{i,j}J_{ij}\mathbf{S}_{i}\cdot\mathbf{S}_{j}
+J_k\sum_{i}\mathbf{S}_{i}\cdot\mathbf{S}_{ic},
\label{model}\end{align}
where the antiferromagnetic Heisenberg strength $J_{ij}$ between local spins $\mathbf{S}_i$ and $\mathbf{S}_j$ equal $J_1$ for N.N. and $J_2$ for N.N.N., respectively. We introduce two cases of $J_2$ configurations, the first is that each spin is connected to six other N.N.N. spins (Fig.\ref{lattice}(a)); the latter is that only one $J_2$ interaction exists in each hexagon (Fig.\ref{lattice}(b)).
The electron (created by $c^\dag_{i\sigma}$) hopping strengths are $t_{ij}=-t$ for N.N. and $t'$ for N.N.N.. Local moments interact with conduction electrons through Kondo coupling $J_k>0$.
The spin-density of itinerant electrons can be written using Pauli matrix as $\mathbf{S}_{ic}=\frac{1}{2}
\sum_{\alpha \beta }c_{i\alpha }^{\dag }\bm{\sigma}_{\alpha \beta }c_{i\beta
}$, while the local spins are represented by slave-fermions as $\mathbf{S}_{i}=\frac{1}{2}\sum_{\alpha
\beta }f_{i\alpha }^{\dag }\bm{\sigma}_{\alpha \beta }f_{i\beta }$ under restriction of only one fermions in each lattice site, which can be composed by adding a Lagrangian term $\sum_{i}\lambda _{i}(\sum_{\sigma
}f_{i\sigma }^{\dag }f_{i\sigma }-1)$ to the Hamiltonian.

\begin{figure}[tbp]
\hspace{-0.1cm} \includegraphics[totalheight=1.8in]{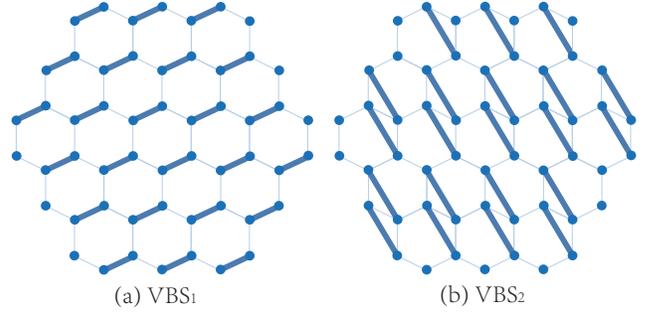}
\caption{(Color online) VBS configurations with lowest energies. (a) VBS$_1$: every two spins connected by $J_1$ form isolated bonds. (b) VBS$_2$: two-spin bonds are created along $J_2$ lines. In these two VBS, the bonds can be set arbitrarily along $J_1$ and $J_2$ lines, respectively, provided that each spin belongs to a single two-spin bond.
}
\label{RVB}
\end{figure}

\begin{figure}[tbp]
\hspace{-0cm} \includegraphics[totalheight=1.75in]{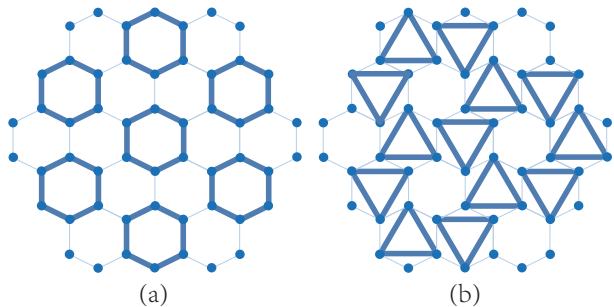}
\caption{(Color online) Two possible VBS considered in our calculations: (a) hexagonal bonds and
(b) triangular bonds. These two states exhibit higher energies than VBS$_1$ and VBS$_2$ in Fig.2.
 }
\label{otherRVB}
\end{figure}

\section{Heisenberg model in slave-fermion representation}
 Before calculating the phase diagram of KHLM, we first study the $J_1$-$J_2$ Heisenberg model on honeycomb lattice in slave-fermion representation. We perform two steps of mean-field decoupling, first we rewrite the Heisenberg interactions to the form (neglecting a trivial constant)~\cite{Liu12}
\begin{align}
J_{ij}\mathbf{S}_{i}\cdot\mathbf{S}_{j}=-\frac{1}{2}\sum_{\sigma\sigma'}f^\dag_{i\sigma}f_{j\sigma}f^\dag_{j\sigma'}f_{i\sigma'},
\end{align}
these terms describe creation of spin-bonds, and are approximated by introducing VBS strength $\chi_{ij}=-\sum_{\sigma}\langle f^\dag_{i\sigma}f_{j\sigma}\rangle$. For the polarization term
$\mathbf{S}^z_{i}\cdot\mathbf{S}^z_{j}$, we introduce the staggered magnetization $m_f=\frac{1}{2}\sum_\sigma\sigma\langle f^\dag_{Ai\sigma}f_{Ai\sigma}\rangle$ ( takes opposite value in B sublattice) to decouple it.
In order to generate more reliable results, the bond-creation term and polarization term are multiplied by weights $\eta$ and $1-\eta$, respectively~\cite{YuRong14}, while $\eta=0.5$ corresponds to common mean-field treatments~\cite{Isaev13}. The mean-field parameters $\lambda,m_f,\chi_{ij}$ are determined by minimizing the ground state energy.

In the VBS phases, different kinds of spin bonds may exist (see Fig.\ref{RVB},\ref{otherRVB}). For example, two-spin bonds (denoted by $\chi_{ij}\ne0$) can be created along either $J_1$ or $J_2$ lines; isolated bonds can be formed by two, three, four, six and even more spins (similar to the dimer patterns in quantum dimer model~\cite{Moessner01}); in addition, all $\chi_{ij}$ along N.N. direction can be nonzero to form unitary VBS. All these VBS should be considered. For VBS consisted of isolated bonds, the energies are easily calculated as summation of the binding energy for each bond. We found the two-spin bond VBS to be the lowest-energy states~\cite{Chayes89}, with $-\frac{1}{4}J_1\eta$ (VBS$_1$ in Fig.\ref{RVB}) and $-\frac{1}{4}J_2\eta$ (VBS$_2$ in Fig.\ref{RVB}) per site. For
hexagonal VBS, triangular VBS (see Fig.\ref{otherRVB}) and the unitary VBS, the energies are calculated to be $-\frac{2}{9}J_1\eta$, $-\frac{1}{8}J_2\eta$ and $-0.2066J_1\eta$ per site, respectively, all higher than the two-spin VBS. For this reason, only VBS$_1$ and VBS$_2$ are included in the phase diagram~\cite{Chayes89}. By comparing their energies, the phase boundary between VBS1 and VBS2 is obtained to be $J_2/J_1=1$, similar to the Heisenberg model on SSL~\cite{YuRong14}. We calculated the energy of AF phase to be $-\frac{3}{4}(J_1-2J_2)(1-\eta)N$, where $N$ is the number of hexagon, therefore the boundary between AF and VBS$_1$ is $J_2/J_1=\frac{3-5\eta}{6(1-\eta)}$.
Furthermore,
AF and VBS may coexist (in SSL model~\cite{YuRong14}), e.g., AF+unitary VBS, AF+hexagonal VBS, etc. However, these states have higher energies than VBS$_1$ phase, therefore they don't appear in the phase diagrams.

\begin{figure}[tbp]
\hspace{-0cm} \includegraphics[totalheight=2.1in]{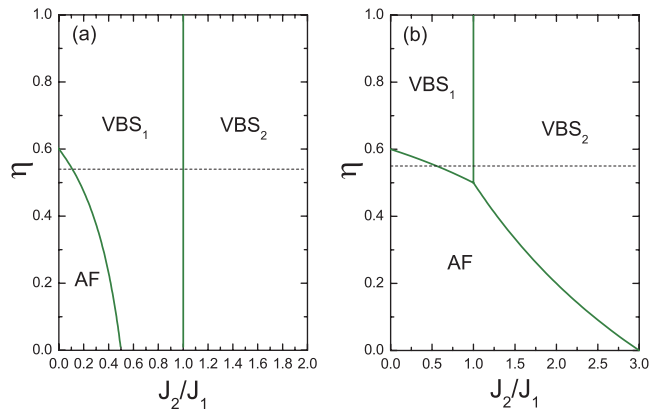}
\caption{(Color online) Phase diagram of $J_1$-$J_2$ Heisenberg model in slave-fermion representation.
(a) and (b) correspond to lattices Fig1.(a) and Fig1.(b), respectively. Dashed lines indicate the choices of $\eta$. For the meaning of $\eta$, see the main text.
 }
\label{Hphasediagram}
\end{figure}

\begin{figure}[tbp]
\hspace{-0.2cm} \includegraphics[totalheight=2.35in]{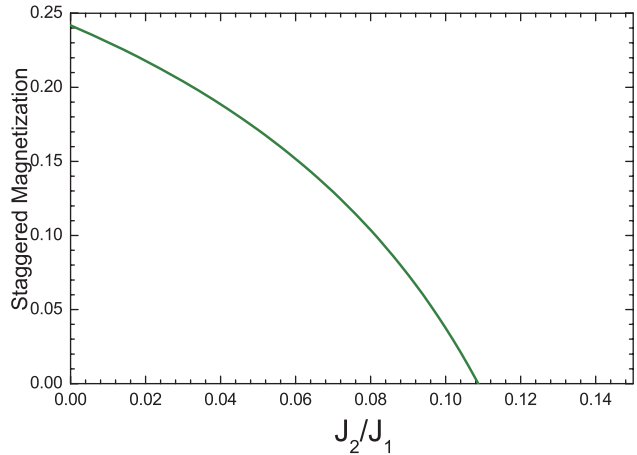}
\caption{(Color online) Spin-wave results for the staggered magnetization of the Heisenberg lattice in Fig1.(a).}
\label{spinwave}
\end{figure}

The phase diagram of Heisenberg lattice in Fig.\ref{lattice}(a) is given in Fig.\ref{Hphasediagram}(a) as a function of frustration $Q=J_2/J_1$ and $\eta$.
AF order is suppressed by enlarged $Q$ rapidly and finally vanishes at a critical frustration $Q=\frac{3-5\eta}{6(1-\eta)}$ depending on the choice of $\eta$.
For comparison, we also calculated the staggered magnetization through linear spin-wave theory, illustrated in Fig.\ref{spinwave}. We found a appropriate choice of $\eta=0.54$ at which AF order vanishes at $Q_c=0.1087$, equal to spin-wave result. In this phase diagram, AF and VBS$_2$ are separated completely by VBS$_1$ phase, this character is owning to the large number of $J_2$ lines, which is twice the number of $J_1$ lines, as a result, the AF order can be easily destroyed by even small $Q$.
Reducing the number of $J_2$ lines may make there phases closer with each other in the phase diagram. For the case of only one $J_2$ line in each hexagon (Fig.\ref{lattice}(b)), the corresponding phase diagram is illustrated in Fig.\ref{Hphasediagram}(b), in which each phase is connected with other two phases. Once the Kondo coupling is included, the phase diagrams of KHLM on Fig.\ref{lattice}(a) and (b) may be distinct from each other. For structure2, we choose $\eta=0.55$, also close to the conventional value $0.5$.

\section{Phase diagram of Kondo-Heisenberg lattice}

The on-site Kondo couplings lead to the screening of local spins by conduction electrons, on the other hand, the longitudinal spin interactions in Kondo couplings may give rise to the polarization of both conduction electrons and local spins in opposite directions. In the mean-field procedure, the Kondo coupling in Eq.\ref{model} is firstly decomposed into singlet and triplet hybridizations between conduction electrons and slave-fermions as~\cite{Isaev13,Li15}
\begin{align}-&\frac{3}{8}J_k\sum_i[(\sum_{\sigma}c^\dag_{i\sigma}f_{i\sigma})\cdot h.c.]\nonumber\\+&\frac{1}{8}J_k\sum_i[(\sum_{\sigma}\sigma c^\dag_{i\sigma}f_{i\sigma})\cdot h.c.].
\end{align}
The first term denotes traditional singlet Kondo screening, which can be estimated by introducing $V_s=\frac{1}{2}\sum_\sigma\langle c^\dag_{i\sigma}f_{i\sigma}\rangle$ to perform Hartree-Fock approximation, while the second term describes triplet paring, which can be approximated by introducing staggered parameter $V_t=\frac{1}{2}\sum_\sigma\sigma\langle c^\dag_{Ai\sigma}f_{Ai\sigma}\rangle$. Note that $V_t$ takes opposite values in different sublattices, and it only survives in the coexisting phase of AF and Kondo screening~\cite{Asadzadeh13,Li15}. The longitudinal Kondo interaction $J_k\sum_i\mathbf{S}^z_i\mathbf{S}^z_{ic}$ can be decoupled by using $m_f$ and $m_c=-\frac{1}{2}\sum_\sigma\sigma\langle c^\dag_{Ai\sigma}c_{Ai\sigma}\rangle$ (both are staggered in different sublattices)~\cite{Zhang00}.

By combining the electron-hopping term together with the Heisenberg and Kondo terms under above mean-field treatments, the Hamiltonian in Eq.\ref{model} can now be expressed in momentum space by matrix form $\mathcal{H}=N\epsilon_0+\sum_{\mathbf{k}\sigma}\Phi^\dag_{\mathbf{k}\sigma}H_{\mathbf{k}\sigma}\Phi_{\mathbf{k}\sigma}$, where $\epsilon_0$ is a constant term and $H_{\mathbf{k}\sigma}$ is a $4\times4$ matrix. The four-component spinon is defined as $\Phi_{\mathbf{k}\sigma}=(c_{A\mathbf{k}\sigma}~c_{B\mathbf{k}\sigma}~f_{A\mathbf{k}\sigma}~f_{B\mathbf{k}\sigma})^T$.
The excitation spectrums are easily obtained by diagonalizing $H_{\mathbf{k}\sigma}$, then the mean-field parameters $V_s$, $V_t$, $m_c$, $m_f$, $\lambda$, $\mu$, $\chi_{ij}$ are determined by minimizing the ground state energy, finally, phases can be  classified according to the non-zero order parameters.
In general case, analytical expressions for the spectrums are unavailable, we have to diagonalize $H_{\mathbf{k}\sigma}$ numerically, through which we obtain the Bogliubov transformations to determine the parameters through their definitions~\cite{Li15}.

\begin{figure}[tbp]
\hspace{-0.2cm} \includegraphics[totalheight=2.5in]{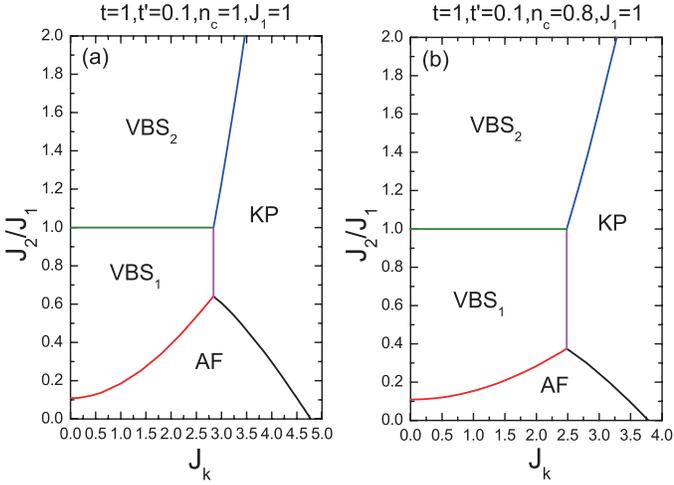}
\caption{(Color online) Phase diagram of Kondo-Heisenberg lattice model on honeycomb lattice (see Fig1.(a)) as a function of Kondo coupling $J_k$ and frustration $J_2/J_1$.
 }
\label{phasediagram1}
\end{figure}

\begin{figure}[tbp]
\hspace{-0.1cm} \includegraphics[totalheight=2.3in]{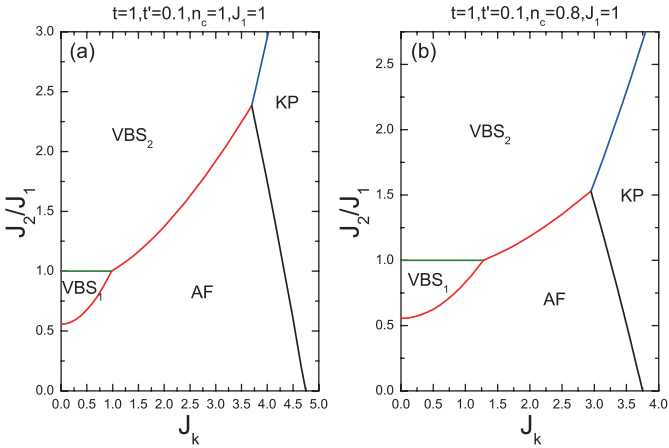}
\caption{(Color online) Phase diagram of Kondo-Heisenberg lattice model on honeycomb lattice (Fig1.(b)) as a function of Kondo coupling and frustration.
 }
\label{phasediagram2}
\end{figure}

The derived phase diagrams of KHLM on honeycomb lattice corresponding to Fig.\ref{lattice}(a) and (b) are shown in Fig.\ref{phasediagram1} and \ref{phasediagram2}, respectively.
For each lattice structure, two cases of electron occupation per site $n_c=1$ and $n_c=0.8$ are considered, causing no essential influence to the phase diagrams.
At $J_k=0$, the phase-evolution with frustration $Q$ follows the
dashed lines in Fig.\ref{Hphasediagram}, from AF phase ($m_f,m_c\ne0$) at small frustration, to VBS$_1$ then to VBS$_2$ phase with increasing frustration.

In weak frustration region ($Q<0.1087$ for structure1, and $Q<0.556$ for structure2), the AF order persists till a critical $J_k$ into KP phase, which is characterized by non-zero Kondo screening strength $V_s$. For the Kondo lattice model on square lattice, previous works based on mean-field approximation, bond-fermion approach, variational Monte Carlo calculations and Gutzwiller approximation have verified the coexistence of
AF and KP at intermediate $J_k$~\cite{Watanabe07,Lanata08, Custers12,Isaev13,Li15,Eder16}, while on the honeycomb lattice, we find this coexistence phase to be higher in energy than both AF and KP phases, therefore it vanishes in the phase diagrams, similar to Kondo insulator defined on SSL~\cite{YuRong15}. This feature may be attributed to the unique density of states (DOS) for conduction electrons. For the electrons hopping on honeycomb lattice, we notice a great reduction of DOS around the Fermi level near half-filling, this small DOS weakens the correlation with local spins, which may result in the instability of coexisting phase between AF and KP.
This result certainly needs further theoretical verification.

With medium frustration($Q<1$) and small $J_k$, the system is in the VBS$_1$ phase, increasing $J_k$ enhances AF correlation between itinerant electrons and local spins, resulting in the recovery of AF order. Further increasing of $J_k$ enhances Kondo screening and
drives AF phase to KP phase. For sturcture1, at $0.64<Q<1$(for $n_c=1$), the VBS$_1$ phase can be shifted directly to KP phase by increased $J_K$. In strong frustration regime($Q>1$), the VBS$_2$ phase enters, then the KP phase arises over a critical $J_k$. Notably, for structure2, we found that over a wide frustration region $0.56<Q<2.385$ (for $n_c=1$), a intermediate $J_k$ can recover the AF order from VBS phases (similar process is also seen in VBS$_1$ phase in $0.1087<Q<0.64$ for structure1). The recovery of AF order from VBS phases by enhancement of Kondo coupling is not realized in square lattice system such as in SSL KHLM~\cite{YuRong14}.

\section{conclusion}

In summary, we have studied the
KHLM on honeycomb lattice with antiferromagnetic N.N.
and N.N.N. Heisenberg interactions $J_1$ and $J_2$, respectively, with the later providing geometrical frustrations.
We employed slave-fermion representation of local spins to study the interplay
between Kondo coupling and magnetic frustration.
We found that in small Kondo coupling region, increased frustration can destroy the magnetic order, driving the
system into disordered VBS phases.
Remarkably, the magnetic transition and breakdown of Kondo screening converges, leaving no coexisting phase of AF and Kondo screening, in contrast to the separation of these two transitions in square lattice systems.
Moreover, we found
two stable VBS phases, in which the spin-bonds are constructed
through N.N. and N.N.N. RKKY exchanges, respectively,
and these two VBS phases are separated by a frustration ratio
$J_2/J_1=1$. Notably, in a rather wide frustration region,
enhancing the Kondo coupling can recover the AF
order form VBS states, further increasing of Kondo coupling leads to KP phase with Kondo screening. Large frustration
strength can destroy the Kondo screening in KP phase and bring
the system into VBS phase. Our result may give
theoretical supports to the experimental exploration
of quantum phase transitions in heavy-fermion compounds
with geometrical frustration.

\acknowledgments

The authors thank Guang-Ming Zhang, Lu Yu and Yi-Feng Yang for earlier collaborations and instructions.
H. Li acknowledges the supports by NSF-China (No. 11504061),
Guangxi NSF (No. 2015GXNSFBA139010) and the Scientific
Research Foundation of Guilin University of technology.
Y. Liu thanks the supports by the China Postdoctoral Science Foundation and the Foundation of LCP. H.F.
Song acknowledges the funding supports from National Natural Science Foundation of China under Grant No. 11176002, National High Technology Research and Development Program of China under Grant 2015AA01A304.

\end{document}